\documentclass[%
 aip,
 amsmath,amssymb,
 preprint,%
]{revtex4-1}

\usepackage{graphicx}
\usepackage{dcolumn}
\usepackage{bm}

\usepackage[utf8]{inputenc}
\usepackage[T1]{fontenc}
\usepackage{mathptmx}
\DeclareMathAlphabet{\mathcal}{OMS}{cmsy}{m}{n}
\usepackage{etoolbox}
\usepackage{color}
\newcommand{\figref}[1]{FIG. \ref{#1}}
\newcommand{\secref}[1]{Section \ref{#1}}
\newcommand{\Alfven}{{Alfv\'en}~}
\newcommand{\Poincare}{{Poincar\'e}~}
\newcommand{\etal}{\textit{et al.}}

\newcommand{\modi}[1]{{\color{black} #1}}
\newcommand{\modii}[1]{{\color{black} #1}}

\makeatletter
\def\@email#1#2{%
 \endgroup
 \patchcmd{\titleblock@produce}
  {\frontmatter@RRAPformat}
  {\frontmatter@RRAPformat{\produce@RRAP{*#1\href{mailto:#2}{#2}}}\frontmatter@RRAPformat}
  {}{}
}%
\makeatother
\begin{document}

\preprint{POP25-AR-CCRD2025-00092}

\title{Sawtooth crash in tokamak as a sequence of Multi-region Relaxed MHD equilibria}
\author{Z. S. Qu}
\affiliation{ 
School of Physical and Mathematical Sciences, Nanyang Technological University,  Singapore 637371, Singapore
}%
 \email{zhisong.qu@ntu.edu.sg}
\author{Y. Zhou}%

\affiliation{ 
School of Physics and Astronomy, Institute of Natural Sciences, and MOE-LSC, Shanghai Jiao Tong University, Shanghai 200240, China
}%



\author{A. Kumar}
\affiliation{%
Plasma Science and Fusion Center, Massachusetts Institute of Technology, Cambridge, MA 02139,
United States of America
}%

\author{J. Doak}
\affiliation{%
Mathematical Sciences Institute, the Australian National University, Canberra ACT 2601, Australia
}%

\author{J. Loizu}
\affiliation{%
\'{E}cole Polytechnique F\'{e}d\'{e}rale de Lausanne, Swiss Plasma Center, CH-1015 Lausanne,
Switzerland
}%

\author{M. J. Hole}
\affiliation{%
Mathematical Sciences Institute, the Australian National University, Canberra ACT 2601, Australia
}%
\affiliation{%
Australian Nuclear Science and Technology Organisation, Locked Bag 2001, Kirrawee DC, NSW 2232, Australia
}%
\date{\today}

\begin{abstract}
This study examines the sawtooth crash phenomenon in tokamak plasmas by modelling it as a sequence of Multi-region Relaxed Magnetohydrodynamic (MRxMHD) equilibria. 
Using the Stepped-Pressure Equilibrium Code (SPEC), we constructed a series of equilibria representing intermediate states during the sawtooth crash, with progressively increasing reconnection regions. 
Numerical results demonstrated that the system prefers the lower energy non-axisymmetric equilibria with islands and is eventually back to an axisymmetric state, capturing key features of the reconnection process. Comparisons with the nonlinear MHD code M3D-C1 showed remarkable agreement on the field-line topology, the safety factor, and the current profile.
However, the simplified MRxMHD model does not resolve the detailed structure of the current sheet. Despite this limitation,  
MRxMHD offers an insightful approach and a complementary perspective to initial-value MHD simulations.
\end{abstract}

\maketitle

\section{Introduction}
Relaxation is a universal phenomenon where plasmas undergo self-organisation to reach a lower energy state~\cite{Taylor1986}.
Modelling this process often involves running initial-value simulations, which can be computationally expensive. Variational principles offer a more efficient alternative by directly constructing the final state through the minimisation of total plasma energy, subject to specific constraints.
A famous example is the Woltjer-Taylor relaxation~\cite{woltjer_theorem_1958,Woltjer1958, taylor_relaxation_1974}, which uses the volume-integrated magnetic helicity as a constraint. This leads to a constant-pressure, constant-current force-free Beltrami field, and a bifurcated non-axisymmetric state for helicity values exceeding a certain threshold.

\begin{figure}[!htbp]
    \centering
    \includegraphics[width=0.48\linewidth]{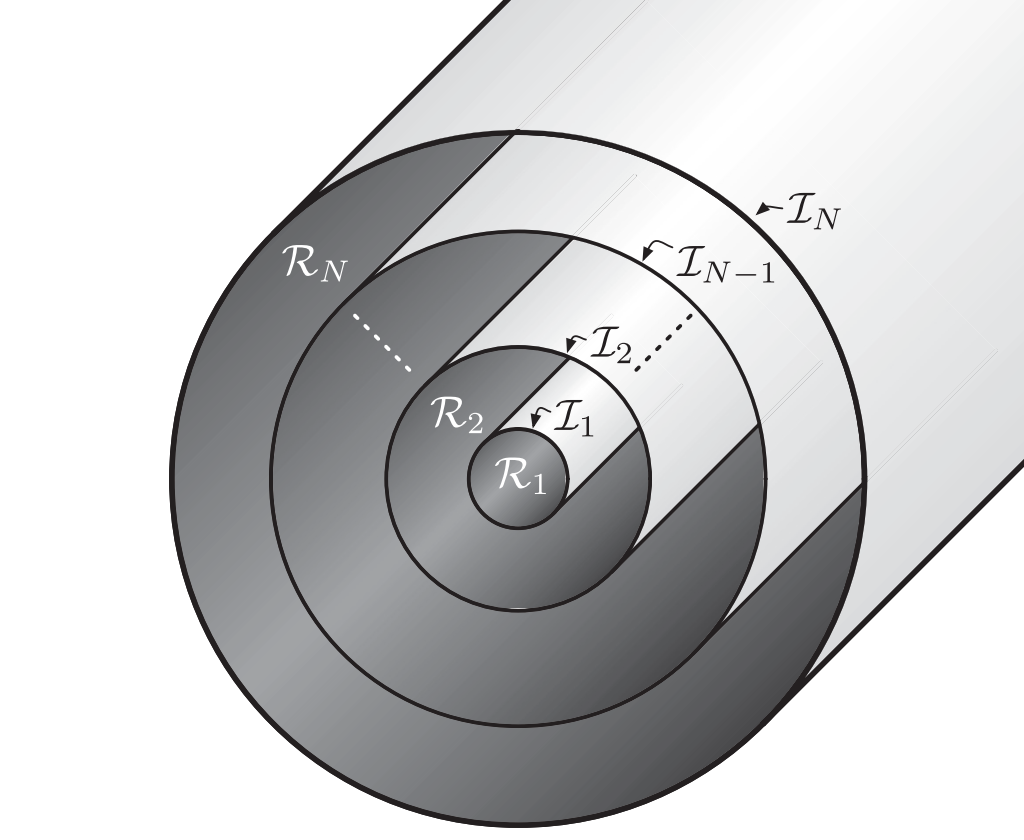} 
    \caption{A schematic view of the plasma regions $\mathcal{R}_i$ and ideal interfaces $\mathcal{I}_i$. Reproduced from [Dennis \etal~ Physics of Plasmas 20(3), 032509 (2013)]\cite{dennis_infinite_2013}, with the permission of AIP Publishing. }
    \label{fig:geometry}
\end{figure}

While the Taylor relaxation successfully modelled early reversed-field pinch (RFP) experiments, where overlapping resistive modes cause global relaxation~\cite{Taylor1986}, such a relaxation is often too drastic for tokamaks and stellarators.
Conversely, the Kruskal-Kulsrud ideal MHD variational principle~\cite{Kruskal1958}, which enforces nested magnetic flux surfaces and prohibits changes in magnetic topology, proves too restrictive for many applications.
To overcome these difficulties, the Multi-region Relax MHD (MRxMHD) was put forward by Robert Dewar based on the Bruno and Lawrence sharp boundary equilibrium~\cite{bruno_existence_1996}.
The model was first published by Hole, Hudson, and Dewar~\cite{hole2006stepped,hole_equilibria_2007,Hole2009}.
The MRxMHD model assumes the plasma volume to consist of $N$ subregions $\mathcal{R}_1, \cdots, \mathcal{R}_N$, separated by $N$ nested ideal interfaces $\mathcal{I}_1, \cdots, \mathcal{I}_{N}$.
In the fixed boundary version, $\mathcal{I}_N$ is a fixed perfectly-conducting wall.
A schematic view of the magnetic geometry is shown in \figref{fig:geometry}.
The magnetic helicity is constrained in each subregion instead of the entire plasma, allowing local relaxations meanwhile preserving the magnetic topology of the prescribed interfaces.
The resulting equilibrium, after minimising the total energy, has a discrete pressure profile with jumps on the interfaces,
namely a stepped-pressure equilibrium.
With a sufficient number of interfaces, the global plasma profiles are recovered discretely.
The Stepped-Pressure Equilibrium Code (SPEC)~\cite{Hudson2012,Qu2020, Hudson2020} is a numerical solver designed to find such equilibria in complicated geometry.

MRxMHD and SPEC have been utilised to study plasma equilibrium and relaxation.
By scanning the location of an ideal interface acting as a transport barrier, a sequence of equilibria was constructed for the RFX-mod, which successfully replicated the transition between the single-helical-axis state and the double-axis state~\cite{dennis_minimally_2013}.
Another good example is the direct prediction of nonlinear tearing mode saturation both in slab~\cite{Loizu2020,balkovic_direct_2025} and in cylindrical~\cite{loizu_nonlinear_2023} geometry,
where the width of the island was recovered without solving an MHD initial value problem.


The main purpose of the current paper is to examine the suitability of MRxMHD in modelling a sawtooth crash \modi{as a sequence of equilibria}. Sawtooth is a well-known relaxation process in tokamaks~\cite{von_goeler_studies_1974} and current-carying stellarators~\cite{nagayama_sawtooth_2003, zanini_eccd-induced_2020}.
It is triggered when a $q=1$ surface presents inside the plasma volume with the temperature profile undergoing periodic crashes and flattenings, leading to worse confinement in general.
Here $q$ denotes the safety factor.
The crash is usually followed by a ramp-up recovering phase during which the plasma internal energy re-accumulates.
The physical mechanism of the sawtooth is still under debate~\cite{chapman_controlling_2011}.
The two most cited models are the Kadomstev model~\cite{kadomtsev1975disruptive} and the Wesson model~\cite{Wesson1986,Wesson1990}.
The Kadomstev model attributes the crash to a $m=n=1$ internal kink mode and magnetic reconnection, 
where $m$ and $n$ are the poloidal and toroidal mode numbers, respectively. 
During a crash, flux surfaces and their neighbourhood with the same helical flux label $\chi=q_{\text{res}} \psi_p  - \psi_t$ reconnects, 
conserving $\chi$ itself and the differential toroidal flux.
Here $\psi_p$ and $\psi_t$ are the poloidal flux and toroidal flux, respectively, while $q_{\text{res}}=1$ is the resonant safety factor.
The Wesson model, on the other hand, relies on interchange instabilities, which are more prominent at a relatively high plasma $\beta$,
where $\beta=2\mu_0p/B^2$ is the ratio between the plasma kinetic energy and magnetic energy,
with $\mu_0$ being the vacuum magnetic permeability, $p$ the plasma pressure, and $B$ the magnetic field strength.

Since variational approaches including MRxMHD contain neither resistivity nor a time variable, it is not purposed to resolve the complicated crash dynamics which depends strongly on the plasma $\beta$ and the non-ideal effects (see the review by Chapman~\cite{chapman_controlling_2011} and references therein).
An application would be to predict the after-crash equilibrium given the initial configuration.
This was pioneered by Bhattacharjee and Dewar~\cite{bhattacharjee_energy_1980,bhattacharjee_energy_1982,bhattacharjee_energy_1983}, 
in which the volume integrated helicity, now weighed by arbitrary functions of $\chi$, is added to the Taylor variational principle as invariants.
Models based on Taylor relaxation in only the plasma core were also considered~\cite{gimblett_calculation_1994}.
Recently, Aleynikova \etal~\cite{aleynikova_model_2021} applied SPEC to compute the after-crash current and rotational transform profile in the Wendelstein 7-X ECCD heating scenario~\cite{zanini_eccd-induced_2020}, 
in which the reconnected region is considered a single subregion after the crash.
SPEC was also used to find the relaxed parallel flow profile at the edge of the Madison Symmetric Torus (MST) before and after a sawtooth crash, showing good agreements with measurements~\cite{Qu2020flow}.

In this work, we take a further step to construct a sequence of MRxMHD equilibria to resolve the intermediate steps during the crash.
We limit our case to a nearly zero $\beta$ plasma so the Kadomstev model is appropriate. 
A sequence-of-equilibria model for sawtooth was first proposed by Waelbroeck\cite{waelbroeck_current_1989,waelbroeck_onset_1993} for crashes slower than the \Alfven time scale, but faster than the resistive time scale,
such that the conservation of helicity is a good assumption.
At every step, an approximate helical equilibrium is found with an inner-layer current sheet matched to the outer region, where each pair of original flux surfaces reconnects into a new crescent flux surface in the island.
Waelbroeck discovered the existence of a ribbon connecting the two Y-points of the island, with a current sheet in the ribbon.
We take a similar approach as Waelbroeck, with a less restrictive relaxation:
the reconnected island is taken as a single Taylor-relaxed subregion without enforcing a constraint on each crescent flux surface, while the rest of the plasma is kept ideal by the enforced interfaces.
As the size of the reconnected region changes, we obtain a sequence of helical states, which have lower energies than the corresponding axisymmetric solutions.
Our approach has the benefit of allowing configurations with a smaller aspect ratio and larger islands, as well as the emergence of chaotic region surrounding the islands which is confirmed by resistive MHD simulations.  

This paper is organised as follows.
\secref{sec:theory} briefly introduces the MRxMHD theory and the SPEC code.
\secref{sec:results} explains the procedure of constructing a sequence of MRxMHD equilibria for a sawtooth crash and presents the numerical results.
A comparison with the nonlinear MHD code M3D-C1 will be presented in \secref{sec:M3DC1}.
Finally, \secref{sec:conclusion} draws the conclusion.

\section{The MRxMHD model}
\label{sec:theory}
We briefly summarised the MRxMHD model~\cite{hole_equilibria_2007,Hole2009} here.
One may refer to Hudson \etal\cite{Hudson2012} and references therein for the mathematical construction of MRxMHD and the steps in the derivations.
We define the MHD energy in the $i$-th region, given by
\begin{equation}
    W_i = \int_{\mathcal{R}_i} \left( \frac{p}{\Gamma-1} + \frac{B^2}{2} \right) dV,
\end{equation}
as well as the magnetic helicity, written as
\begin{equation}
    K_i = \frac{1}{2} \int_{\mathcal{R}_i} \bm{A} \cdot \bm{B} dV,
\end{equation}
where $\Gamma$ is the adiabatic index, $V$ the volume, $\bm{A}$ the magnetic vector potential, $\bm{B} = \nabla \times \bm{A}$ the magnetic field, and $B=|\bm{B}|$
(equivalent to $B /\sqrt{\mu_0}$ in SI unit).
To find an equilibrium state, we seek to extremise the MRxMHD functional given by
\begin{equation}
    F = \sum_{i=1}^N W_i - \mu_i (K_i - K_{0,i}),
    \label{eq:MRxMHD_function}
\end{equation}
where $\mu_i$ is the Lagrange multiplier, and $K_{0,i}$ the constant helicity constraint in each subregion.
A few additional constraints are to be noted: (1) $p_i V_i^\Gamma = S_i$ in each subregion, where $V_i$ is the volume of each subregion and $S_i$ a constant; (2) both the toroidal magnetic flux $\Delta\psi_{t,i}$ and the poloidal flux $\Delta \psi_{p,i}$ in each subregion are constrained to constants $\Delta\psi_{t0,i}$ and $\Delta\psi_{p0,i}$, respectively (the innermost subregion only needs the toroidal flux constraint); (3) the magnetic field is tangential to the interfaces, i.e. the interfaces are ideal MHD barriers.

Extremising \eqref{eq:MRxMHD_function} with respect to $\bm{A}$ leads the Beltrami field equation in each subregion, given by
\begin{equation}
    \nabla \times \bm{B} = \mu_i \bm{B}, \quad \text{within } \mathcal{R}_i,
    \label{eq:Beltrami}
\end{equation}
with the tangential boundary condition mentioned above given by
\begin{equation}
    \bm{B} \cdot \hat{\bm{n}} = 0, \quad \text{on } \mathcal{I}_i,
    \label{eq:ideal_boundary}
\end{equation}
where ${\hat{\bm{n}}}$ is the normal unit vector on the interfaces.
In the view of the variational principle presented here, $\mu_i$ is a dependent variable determined by matching the helicity constraint.
Extremising \eqref{eq:MRxMHD_function} with respect to the location of $\mathcal{I}_i$ leads to the force balance condition on each interface given by
\begin{equation}
    \left[\left[ p + \frac{B^2}{2} \right]\right] = 0, \quad \text{on } \mathcal{I}_i,
    \label{eq:force}
\end{equation}
where $\left[\left[ \cdots \right]\right]$ stands for the difference across the interface.
The pressure is then a constant in each subregion, determined by the constraint $p_i V_i^\Gamma = S_i$.
We emphasise that one needs to specify four constraints for each subregion: $\Delta\psi_{t0,i}$, $\Delta \psi_{p0,i}$, $K_{0,i}$, and $p_i$ (or $S_i$). Again, for the innermost subregion, the constraint $\Delta \psi_{p0,i}$ is dropped.

The SPEC code solves the MRxMHD equilibrium problem numerically for both tokamaks and stellarators.
The vector potential in each subregion is discretised by Chebyshev (Zernike for the innermost volume) polynomials in the radial coordinate $s$ and Fourier series in the generalised poloidal and toroidal angles $\vartheta$ and $\zeta$. The interface locations $R(\vartheta, \zeta)$ and $Z(\vartheta, \zeta)$ are also given in Fourier series, where $R$ and $Z$ are points on the interface in cylindrical coordinates. 
A Beltrami field solver solves \eqref{eq:Beltrami} and \eqref{eq:ideal_boundary} with an initial guess of the interface location, giving $\bm{A}$ as output.
The force mismatch on interfaces is then computed from $\bm{A}$ and the interfaces are moved using Newton's method to satisfy the force balance condition \eqref{eq:force}. An additional spectral condensation constraint\cite{hirshman_optimized_1985} is added to ensure the uniqueness of the boundary representation.
The iteration continues until a tolerance is reached.

\section{Results}
\label{sec:results}
\subsection{Equilibrium setup}\label{eqsetup}
\begin{figure}[!htbp]
    \centering
    \includegraphics[width=0.48\linewidth]{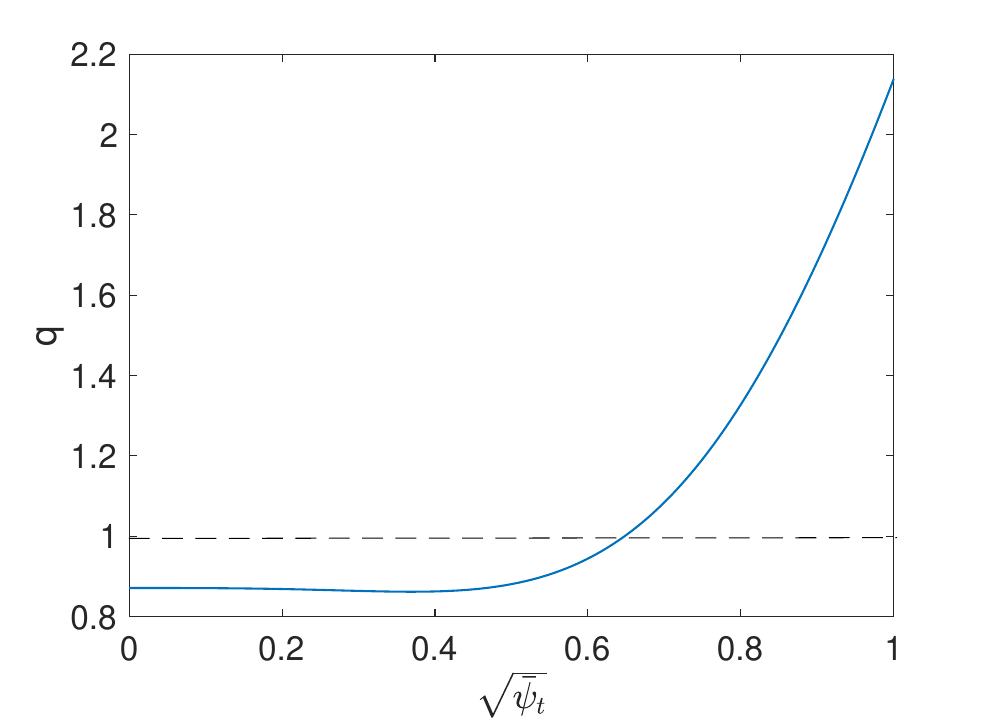} 
    \caption{The $q$ profile of the initial equilibrium.}
    \label{fig:qprof}
\end{figure}

We start from a tokamak plasma with zero pressure, major radius $R_0=1m$, minor radius $a=0.3m$, vacuum field strength $B_0=1T$, and a circular cross-section.
The $q$ profile is chosen to be flat in the core region and increase to towards $2.1$ at the edge, as shown in \figref{fig:qprof}.
This $q$ profile is taken from the class of analytical equilibria in M3D-C1 used to study sawtooth regularly~\cite{M3DC1_manual}.
There is a $q=1$ surface at mid-radius $\bar{\psi}_t=0.42$, or equivalently $\sqrt{\bar{\psi}_t}=0.648$, where $\bar{\psi}_t=\psi_t/\psi_{t,edge}$ is the normalised toroidal flux.
The equilibrium is constructed in VMEC~\cite{Hirshman1983} with a fixed circular boundary.

\begin{figure}[!htbp]
    \centering
    \includegraphics[width=0.48\linewidth]{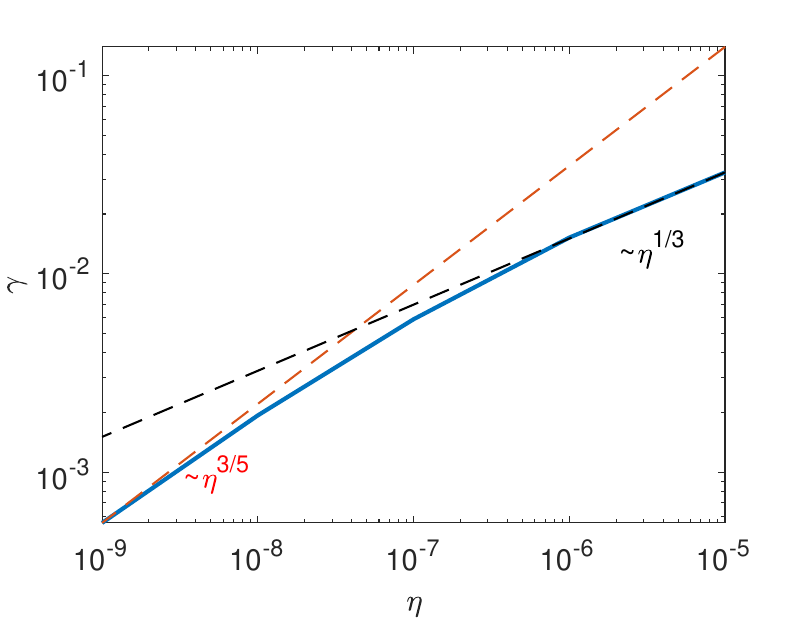} 
    \caption{Resistive instability growth rate calculated by PHOENIX, showing a transition between a tearing mode ($\gamma \sim \eta^{3/5}$) to a resistive-kink ($\gamma \sim \eta^{1/3})$ as the plasma resistivity increases.
    }
    \label{fig:gamma}
\end{figure}

A linear resistive MHD stability calculation with PHONIEX~\cite{blokland_phoenix_2007} shows the equilibrium is unstable to tearing/resistive-kink mode depending on the resistivity.
This is demonstrated in \figref{fig:gamma} where the linear growth rate $\gamma$ is plotted as a function of plasma resistivity $\eta$.
The growth rate $\gamma$ is normalised to the \Alfven frequency on axis $\omega_{A0}=V_{A0}/R_0=B_0/(R_0 \sqrt{\mu_0 \rho_0})$, where $\rho_0$ is the density on axis and $\mu_0$ the plasma vacuum permeability,
while the resistivity $\eta$ is in the unit of $\mu_0 \omega_{A0} R_0^2$, i.e. the inverse \modii{Lundquist} number.
At low resistivity, the growth rate follows the scaling $\gamma \sim \eta^{3/5}$, a typical relationship for a tearing mode. At a higher resistivity, it gradually transits into a resistive kink mode~\cite{coppi1976resistive} with scaling $\gamma \sim \eta^{1/3}$. 
Note that the equilibrium is stable to the ideal internal kink mode~\cite{Bussac1975} as no instability is found in the limit $\eta \rightarrow 0$ since $\beta=0$, and the cross-section circular.

\subsection{Initial MRxMHD equilibrium}
After setting up the ideal MHD equilibrium in VMEC, we need to convert it into a corresponding MRxMHD equilibrium after partitioning into a number of subregions. 
In axisymmetric geometry, one should use an infinite number of subregions to approximate closely a continuous pressure/$q$ profile.
In practice, the number of subregions is finite to ensure numerical viability.
In this work, we start with 15 subregions.

The vector potential and magnetic field in VMEC are written as
\begin{equation}
    \bm{A} = \psi_t \nabla \theta^* - \psi_p \nabla \zeta,
    \label{eq:VMEC_A}
\end{equation}
\begin{equation}
    \mathbf{B} = \nabla \zeta \times \nabla \psi_p + \nabla \psi_t \times \nabla \theta^*,
    \label{eq:VMEC_B}
\end{equation}
where $\psi_t$ labels the flux surfaces and $\psi_p$ is a function of $\psi_t$, $\zeta$ the cylindrical angle, and $\theta^*$ the straight-field-line angle in PEST coordinates.
The helical flux can be computed straightforwardly for each flux surface as $\chi(\psi_t)=\psi_p-\psi_t$.
First of all, it is reasonable to place an ideal interface to separate the region participating or not in the sawtooth reconnection event.
According to the Kadomtsev model, this interface should be placed at the flux surface with a zero helical flux $\chi$.
Flux surfaces outside the $\chi=0$ surface will have a negative value of helical flux, so there is no corresponding flux surface in the core to reconnect with.
Then, we choose to place another interface halfway between this surface and the plasma boundary in the non-reconnecting region to better match the VMEC solution and to account for the changing current profile.

\begin{figure}[!htbp]
    \centering
    \includegraphics[width=0.48\linewidth]{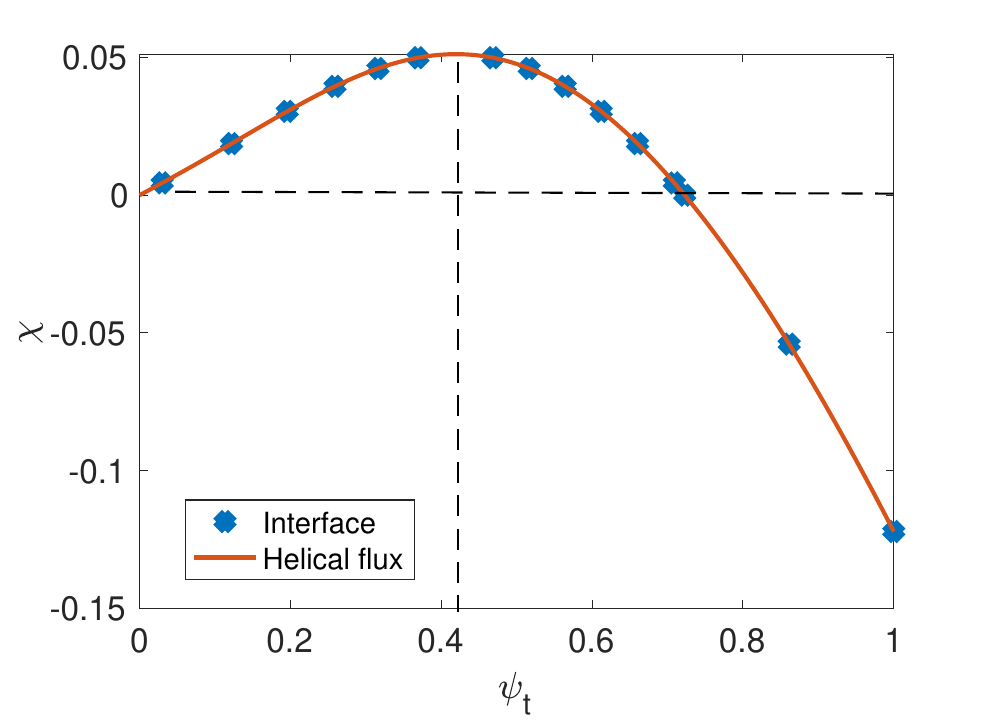}
    \caption{The helical flux used to subdivide the plasma volume and the location of the interfaces (indicated by the blue diamonds.
    The vertical dashed line stands for the $q=1$ surface and the horizontal dashed line stands for $\chi=0$.}
    \label{fig:initial_setup}
\end{figure}

Within the reconnection region, we also need to set up a few interfaces.
Again, according to the Kadomtsev model, flux surfaces with the same helical flux will reconnect.
We therefore place pairs of interfaces at equal helical fluxes based on the VMEC equilibrium.
This is demonstrated by \figref{fig:initial_setup}.
Starting from the subregion containing the $q=1$ surface, we put $\mathcal{I}_{-1}$ inside the $q=1$ surface and $\mathcal{I}_{1}$ outside, such that $\chi(\mathcal{I}_{-1})=\chi(\mathcal{I}_{1})$.
Similarly, we place $\mathcal{I}_{-2}$ and $\mathcal{I}_{2}$, $\mathcal{I}_{-3}$ and $\mathcal{I}_{3}$, $\cdots$, $\mathcal{I}_{-6}$ and $\mathcal{I}_{6}$, until we are sufficiently close to the non-reconnecting interface at $\chi=0$.
The distance between each pair of interfaces is equidistant in $\sqrt{\chi}$,
so the width of each subregion is similar.
\modi{We note that the final result is not very sensitive to the number of interfaces, as long as they are densely packed in the yet-to-reconnect region and the pairs are chosen to have equal helical flux.}
Once we have determined the location of the interfaces in terms of VMEC flux label $\psi_t$, we can compute the constraints for each SPEC subregion from VMEC output.
The constraints $\Delta\psi_t$ and $\Delta\psi_p$ are computed by taking the difference between their corresponding VMEC values on the neighbouring interfaces.
Finally, we need to compute the helicity in each volume.
Using \eqref{eq:VMEC_A} and \eqref{eq:VMEC_B}, the helicity in VMEC between two flux surfaces $\psi_{t,1}$ and $\psi_{t,2}$ can be computed as
\modi{
\begin{equation}
    K_{\text{VMEC}} =2\pi^2 \int_{\psi_{t,1}}^{\psi_{t,2}} \left(\frac{\psi_t}{q} - \psi_p\right) d\psi_t, 
\end{equation}
}
where $\psi_p'(\psi_t)=1/q$ has been used.
\modi{Considering the gauge} convention for SPEC~\cite{Hudson2012}, the relationship between VMEC helicity and SPEC helicity is given by
\modi{
\begin{equation}
    K_{\text{SPEC}} = \left\{
    \begin{array}{cc}
    K_{\text{VMEC}} + 2\pi^2\psi_{t,>} \psi_{p,>}             &  \text{toroidal region}\\
    K_{\text{VMEC}} - 2\pi^2(\Delta\psi_p \psi_{t,<} + \Delta\psi_t \psi_{p,<})    &  \text{annulus region}
    \end{array}
    \right. ,
\end{equation}
}
where $\psi_{t,<}$ and $\psi_{p,<}$ are the corresponding fluxes at the inner interface,
while  $\psi_{t,>}$ and $\psi_{p,>}$ are \modi{those at the outer interface.
We label the enclosed flux by $\Delta \psi_t = \psi_{t,>} - \psi_{t,<}$ and $\Delta \psi_p = \psi_{p,>} - \psi_{p,<}$.}
During the ``reconnection'', we will remove the interfaces in pairs to reconnect flux surfaces with the same helical flux.
After an interface is removed, the constraints for the new combined subregion are recomputed from VMEC given the new interfaces.

\begin{figure}[!htbp]
    \centering
    \includegraphics[width=0.48\linewidth]{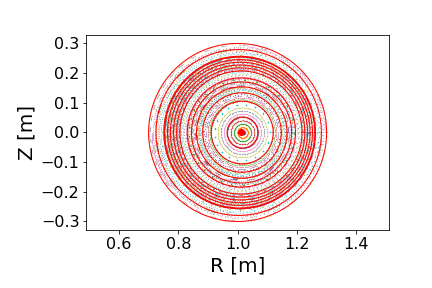}
    \caption{The \Poincare map for the initial SPEC equilibrium with 15 subregions. The interfaces are represented by the red solid lines.}
    \label{fig:initial_spec}
\end{figure}

Given the selected interfaces, the toroidal flux, the poloidal flux, and the magnetic helicity for each subregion are computed from the VMEC magnetic field. 
They are prescribed as constraints for SPEC.
The initial locations of the interfaces are also extracted from VMEC solution.
SPEC then solves the force balance by adjusting the location of the interfaces, assuming axisymmetry.
The resulting equilibrium cross-section is presented in \figref{fig:initial_spec}.
Here we use a poloidal Fourier resolution of $M_{\text{pol}}=10$ and a toroidal resolution $N_{\text{tor}}=7$.
The innermost region uses Zernike polynomials with order 14 while the other regions use Chebyshev polynomials with order 6.

\subsection{A sequence of equilibria representing the crash process}
\begin{figure*}[!htbp]
    \centering
    \begin{tabular}{c c}
      \includegraphics[width=0.48\linewidth]{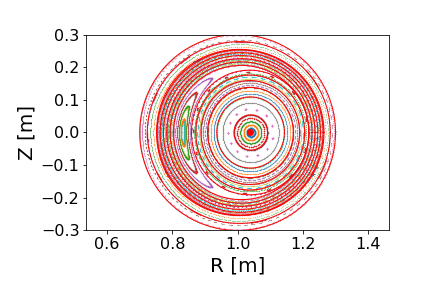}   &  
      \includegraphics[width=0.48\linewidth]{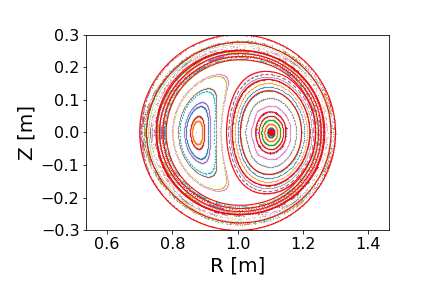}\\
      (a)   &  (b) \\
      \includegraphics[width=0.48\linewidth]{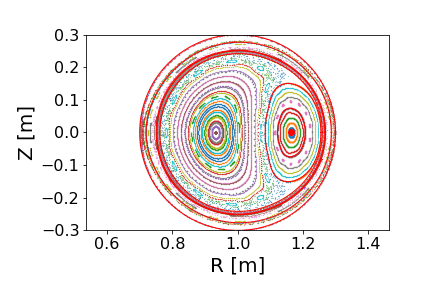}   &  
      \includegraphics[width=0.48\linewidth]{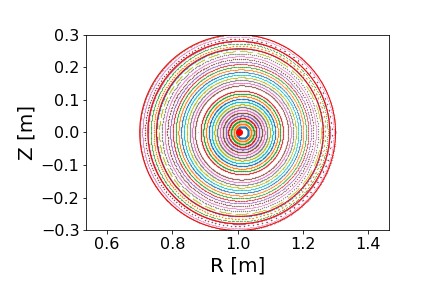}\\
      (c)   &  (d)
    \end{tabular}
    \caption{\Poincare maps of the lowest energy states as we combine the subregions: (a) 15 subregions; (b) 11 subregions; (c) 7 subregions; (d) 3 subregions. The interfaces are represented by the red solid lines.}
    \label{fig:bifurcated}
\end{figure*}

Since the plasma is unstable, the axisymmetric solution is not in the lowest energy state. This is confirmed by calculating the Hessian matrix~\cite{kumar_computation_2021,kumar_nature_2022} of the energy functional for the initial equilibrium, which turns out to have negative eigenvalues, indicating a direction of perturbation that can reduce the system's energy.
To reach the lowest energy state, we first apply an $m=1, n=1$ perturbation to all the interfaces to break symmetry.
Now the configuration no longer has force balance.
A force descent algorithm is then used to push all the interfaces in the direction of the force, i.e. towards the lower $B^2/2$ side across the interfaces.
During the force descent, the interfaces inside the $q=1$ interfaces are continuously pushed into the outer interfaces.
We stop the descent before they crash into each other, and then switch back to a Newton method.
Since we are already very far from the initial equilibrium, the Newton method will not give us the axisymmetric configuration, but rather, a new equilibrium state with a helical core: this is a lower energy state that the system will prefer.
This equilibrium is shown in \figref{fig:bifurcated} (a).
To ensure the existence and robustness of the solution, one should in principle increase the Fourier resolution and study the convergence~\cite{qu_non-existence_2021}.
However, an increased Fourier resolution will result in a smaller gap between the two interfaces, leading to a singular Jacobian and a poorer convergence property or even a lack of convergence.
Given this numerical difficulty, the Fourier resolution is set to be constant in our current paper.

An inspection of \figref{fig:bifurcated} (a) shows that a magnetic island is created, indicating that the flux surfaces between the two interfaces $\mathcal{I}_{-1}$ and $\mathcal{I}_{1}$ are already reconnected. More importantly, these two interfaces almost touch each other on the opposite side of the island and form a very thin layer between them: this is an island with two Y-points connected by a ribbon~\cite{waelbroeck_current_1989, aydemir_nonlinear_1992}.
We will analysis the properties of the current sheet in \secref{sec:M3DC1}.
The next step is to remove $\mathcal{I}_{-1}$ and $\mathcal{I}_{1}$ assuming they are reconnected, and merge the neighbouring subregions.
We simply just add together their fluxes and helicity of the three subregions (the original subregion containing $q=1$ and its neighbours) to form a new subregion.
Now the reconnection site is surrounded by $\mathcal{I}_{-2}$ and $\mathcal{I}_{2}$.
After removing a pair of interfaces, the equilibrium is recomputed.

Starting from the initial 15-subregion equilibrium, we remove 2, 4, and 6 pairs of interfaces, leading to an 11-subregion, a 7-subregion, and a 3-subregion equilibrium, respectively. 
We show the \Poincare maps of this sequence in \figref{fig:bifurcated}.
As more and more subregions/interfaces are reconnected,
the magnetic island is growing bigger,
while the subregion containing the original magnetic axis is becoming smaller.
When a larger number of subregions are reconnected in \figref{fig:bifurcated} (c), a chaotic region is developed around the island's separatrix.
Finally, after all the subregions are reconnected in \figref{fig:bifurcated} (d), the minimum energy state restores to an axisymmetric state, and the magnetic axis is replaced by the island axis.

\begin{figure}[!htbp]
    \centering
    \includegraphics[width=0.48\linewidth]{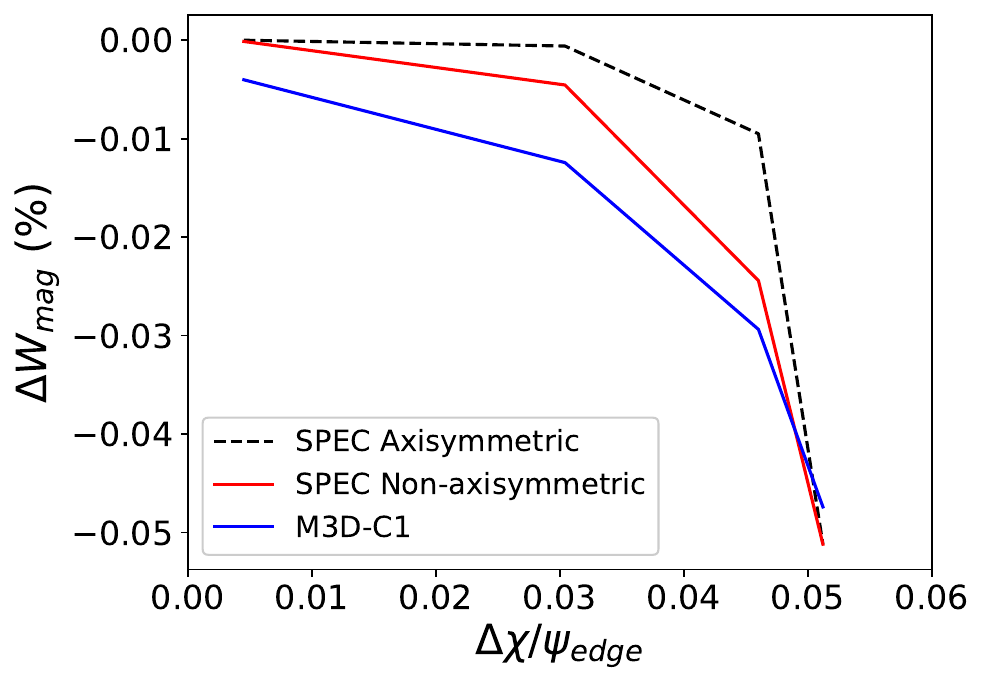}
    \caption{Total change in total magnetic energy compared to the reference case as a function of reconnected helical flux. For SPEC, the reference is the axisymmetric 15-subregion equilibrium. For M3D-C1, the reference is the time slice $t=1350$ just before the island is visible. }
    \label{fig:energy}
\end{figure}

In \figref{fig:energy}, we plot the total change of magnetic energy $\int dVB^2/2$ of the entire plasma as a function of reconnected helical flux $\Delta \chi$, for both the axisymmetric solution and the non-axisymmetric equilibrium as a percentage of the axisymmetric 15-subregion equilibrium.
The non-axisymmetric equilibrium has a lower MHD energy than the corresponding axisymmetric equilibrium.
The difference in energy is very small (on the order of 0.01\%), even though the magnetic topology changes dramatically, consistent with the finding of Cooper \etal~\cite{cooper2010tokamak}. 
It is noteworthy that even when the equilibrium remains axisymmetric as we combine subregions, the total energy will drop.
This is because we are reducing the number of constraints (interfaces) which will naturally lead to a lower total energy.

To sum up, the sequence of MRxMHD equilibria we constructed by removing ideal interfaces represents the process of a sawtooth reconnection.
These are bifurcated states with a lower MHD energy compared to their axisymmetric counterparts.


\section{Comparison with M3D-C1}
\label{sec:M3DC1}

\begin{figure*}[!htbp]
    \centering
    \begin{tabular}{c c}
      \includegraphics[width=0.48\linewidth]{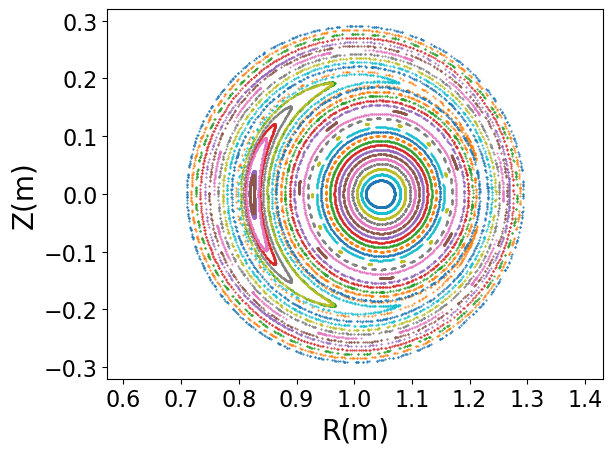}   &  
      \includegraphics[width=0.48\linewidth]{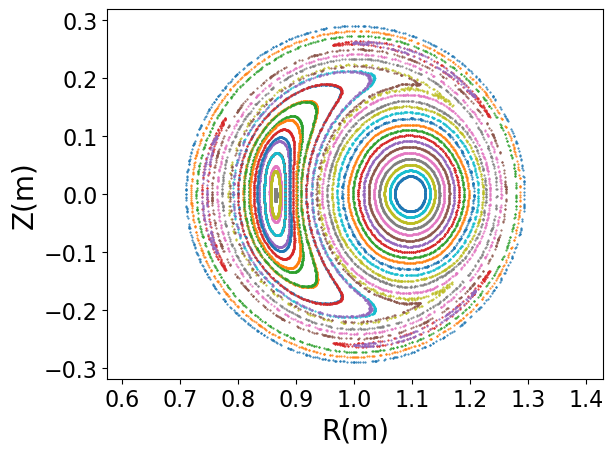}\\
      (a)   &  (b) \\
      \includegraphics[width=0.48\linewidth]{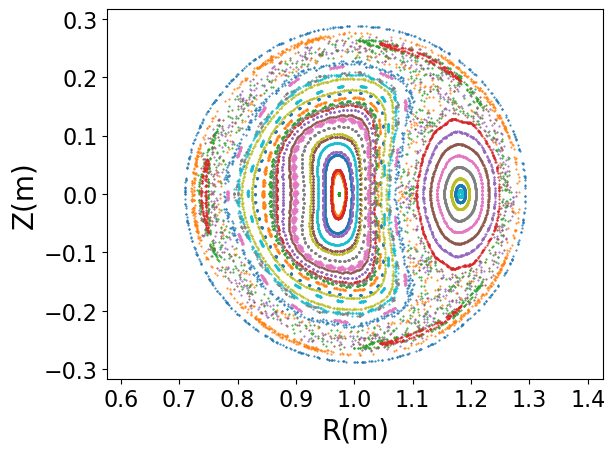}   &  
      \includegraphics[width=0.48\linewidth]{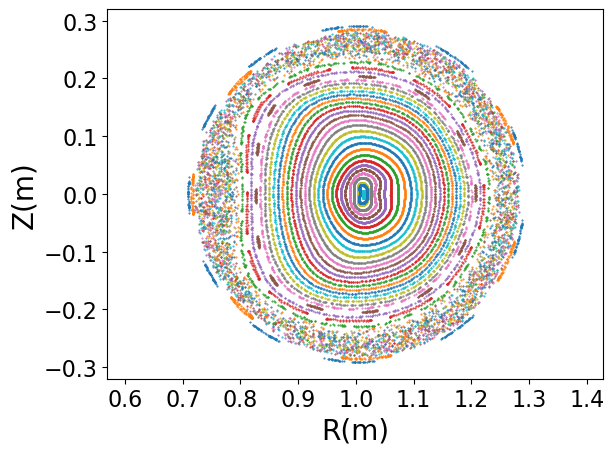}\\
      (c)   &  (d)
    \end{tabular}
    \caption{\Poincare maps of the four time slices of the M3D-C1 simulation that are close to the four SPEC equilibria: (a) $t=1450$; (b) $t=1550$; (c) $t=1650$; (d) $t=1750$.}
    \label{fig:poincare_m3dc1}
\end{figure*}

We now compare the sequence we obtained in \secref{sec:results} to the simulation result of M3D-C1, an initial-value MHD code.
M3D-C1~\cite{jardin_multiple_2012} solves the nonlinear extended MHD equations using high-order finite elements with $C^1$ continuity and split-implicit time-stepping.
The M3D-C1 has long been used to study sawtooth crash and sawtooth cycles in tokamaks~\cite{jardin_multiple_2012, jardin_self-organized_2015, jardin_new_2020} and stellarators~\cite{zhou_nonlinear_2023}.
In this work we use the same VMEC equilibrium to initiate the stellarator extension of M3D-C1~\cite{zhou_approach_2021} to ensure the consistency of the equilibrium.

Specifically, we solve the single-fluid extended MHD equations (in dimensionless units, whose normalisations are consistent with those in \secref{sec:results}. Time is normalised to $\tau_A=\omega_{A0}^{-1}$, length to $R_0$, pressure to $B_0^2$, density to $\rho_0$, mass to proton mass)
\begin{align}
\partial_t \rho + \nabla\cdot(\rho\bm{v}) &= D\nabla^2(\rho-\rho_\text{eq}),\label{continuity}\\
\rho(\partial_t \bm{v} + \bm{v}\cdot\nabla\bm{v}) &= \bm{j}\times\bm{B} - \nabla p - \nabla\cdot\bm{\Pi},\label{momentum}\\
\partial_t p + \bm{v}\cdot\nabla p +\Gamma p\nabla\cdot\bm{v} &= (\Gamma-1)[\eta ({j}^2-{j}_\text{eq}^2) - \nabla\cdot\bm{q} - \bm{\Pi} : \nabla\bm{v}],\label{energy}\\
\partial_t \bm{B} &=  \nabla\times[\bm{v}\times\bm{B} - \eta (\bm{j}-\bm{j}_\text{eq})],\label{induction}
\end{align}
for the mass density $\rho$, velocity $\bm{v}$, pressure $p$, and magnetic field $\bm{B}$, with the current density $\bm{j}=\nabla\times\bm{B}$ and the adiabatic index $\Gamma=5/3$.
The viscous stress tensor is $\bm{\Pi} = -\nu(\nabla\bm{v}+\nabla\bm{v}^{\text{T}}) - 2(\nu_{\text{c}}-\nu)(\nabla\cdot\bm{v})\bm{I}$ and the heat flux 
$\bm{q} = -\kappa_\perp\nabla(T-T_\text{eq}) - \kappa_\parallel\bm{b}\bm{b}\cdot\nabla T$
with $\bm{b} =\bm{B}/B$ and the temperature $T=M p/\rho$, where $M$ is the ion mass. The transport coefficients used include uniform mass diffusivity $D=10^{-6}$, isotropic and compressible viscosities $\nu=\nu_{\text{c}}=10^{-6}$ (in the unity of inverse Reynolds number), resistivity $\eta = 10^{-6}$ (in the unit of inverse \modii{Lundquist} number) and strongly anisotropic parallel and perpendicular thermal conductivities $\kappa_{\parallel}=1$ and $\kappa_{\perp}=10^{-6}$.
For this value of $\eta$, the relationship between $\gamma$ and $\eta$ follows $\gamma \sim \eta^{1/3}$ as shown in \figref{fig:gamma}, and therefore the linear instability is dominated by the resistive kink mode.
The equilibrium fields $\rho_\text{eq}$, $T_\text{eq}$, and $j_\text{eq}$ subtracted in the dissipative terms act as effective sources to sustain the equilibrium in the absence of instabilities, and we consider uniform equilibrium density ($\rho_\text{eq}=1$) and pressure ($p_\text{eq}=10^{-3}$) profiles for simplicity.
We use 3807 reduced quintic elements in the $(R,Z)$ plane and 8 Hermite cubic elements in the toroidal direction, and the time step size is $2$. The boundary conditions are ideal on the magnetic field, no-slip on the velocity, and Dirichlet on the density and pressure. 

We have selected four different time slices (a) $t=1450$, (b) $t=1550$, (c) $t=1650$, and (d) $t=1750$ (in the unit of $\tau_A$).
\modi{The size of the reconnection region in these four slices is visually close} to the four different SPEC equilibria in \figref{fig:bifurcated}, namely the 15-, 11- 7-, and 3-subregion equilibria, respectively. The corresponding \Poincare maps are given in \figref{fig:poincare_m3dc1}.
The toroidal plane shown is chosen to match the island phasing in SPEC.
They are very close to the SPEC equilibria presented in \figref{fig:bifurcated} by visual comparison.
The duration of the crash is $300\tau_A$,
justifying the assumption that the crash is much slower than the \Alfven time scale $\tau_A$ but faster than the resistive time scale $\tau_R \sim \eta^{-1}  = 10^6 $.
The early phase of the crash in \figref{fig:bifurcated} and \figref{fig:poincare_m3dc1} (a) and (b) are almost identical.
This is because in the early stage, the growth of the island is still slow and the reconnection layer is thin, such that an equilibrium model is more appropriate.
There are more notable differences in the later phase of the crash in figures (c) and (d).
In SPEC equilibrium (c), the chaotic region only appears around the good flux surfaces surrounding the original and the new magnetic axis, and is bounded by the non-reconnecting interface.
This is due to our assumption that the crash only affects plasma within this interface.
However, in M3D-C1 the crash is more dramatic and the chaotic region extends beyond the non-reconnecting flux surface to the edge.
\modi{
In future work, we can remove this interface to allow more relaxation, which might give a better match to M3D-C1.}
Finally, SPEC predicts an axisymmetric after-crash configuration,
while in M3D-C1, the outer region remains chaotic for a very long time.

\begin{figure*}[!htbp]
    \centering
    \begin{tabular}{c c}
    \includegraphics[width=0.48\linewidth]{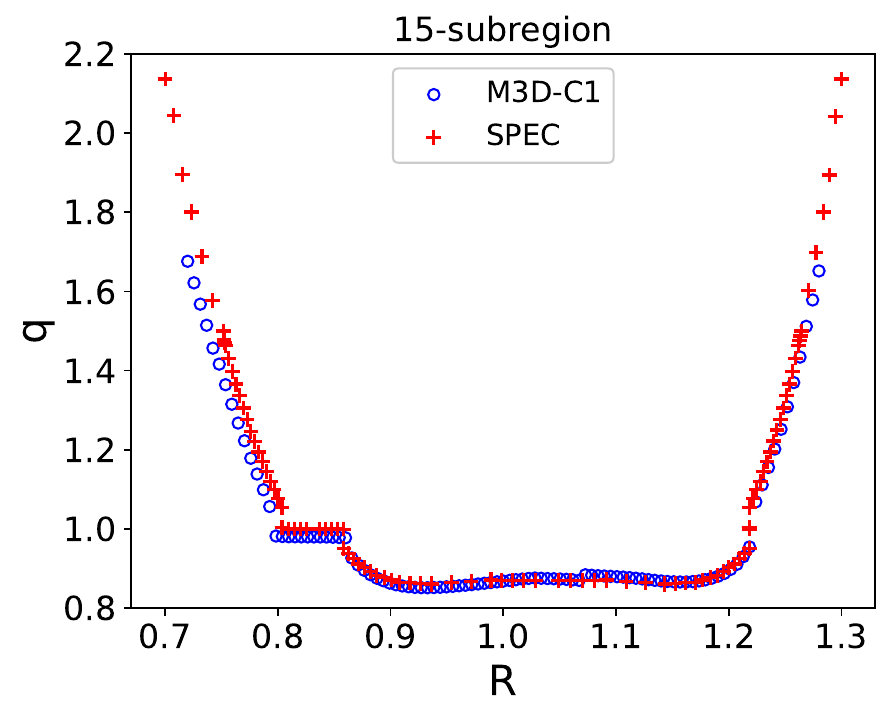}
    &
    \includegraphics[width=0.48\linewidth]{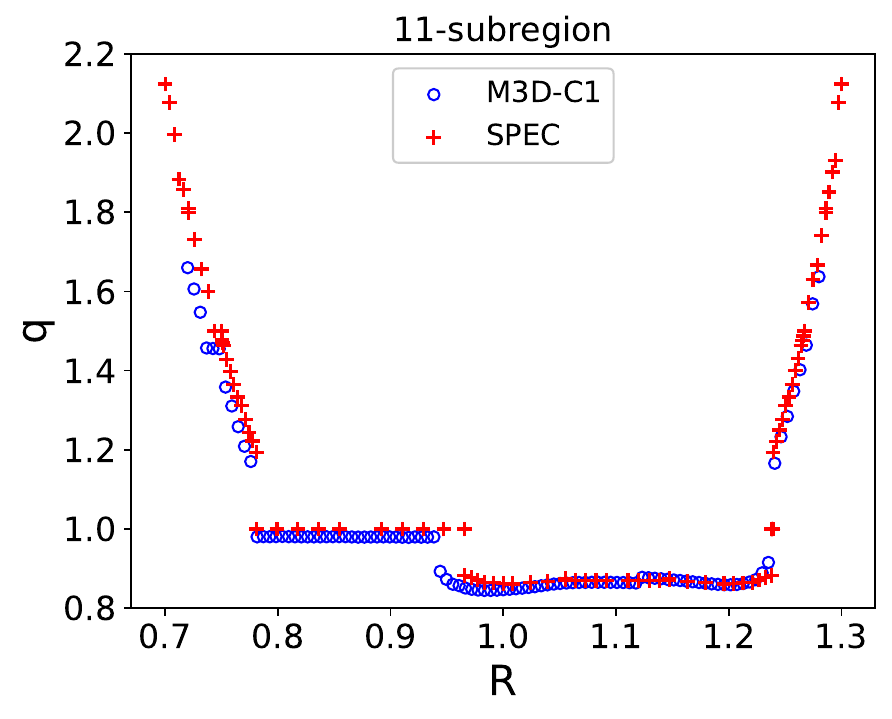}
    \\
    (a) & (b) \\
    \includegraphics[width=0.48\linewidth]{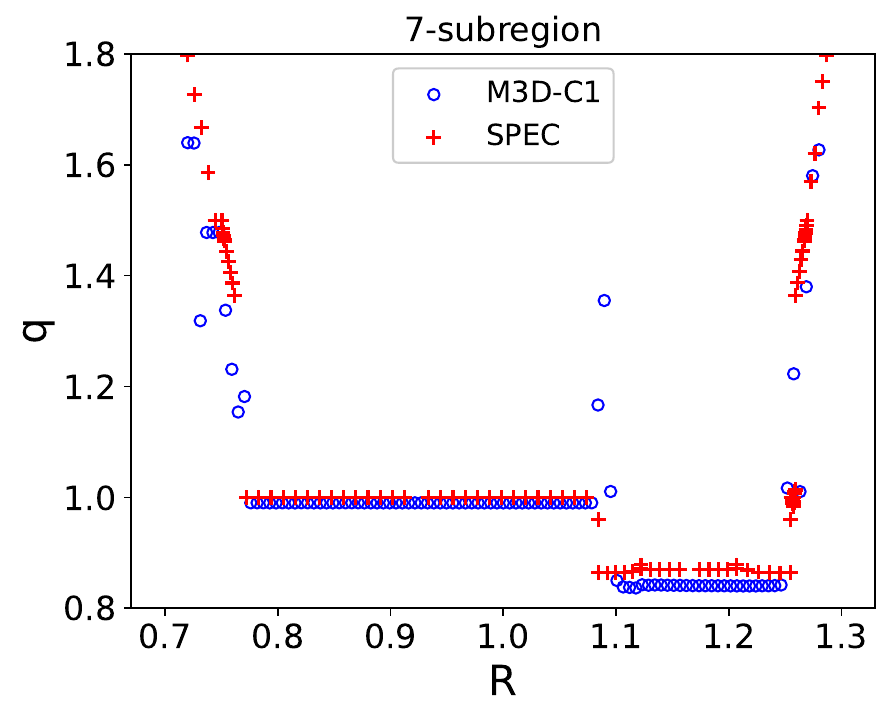}
    &
    \includegraphics[width=0.48\linewidth]{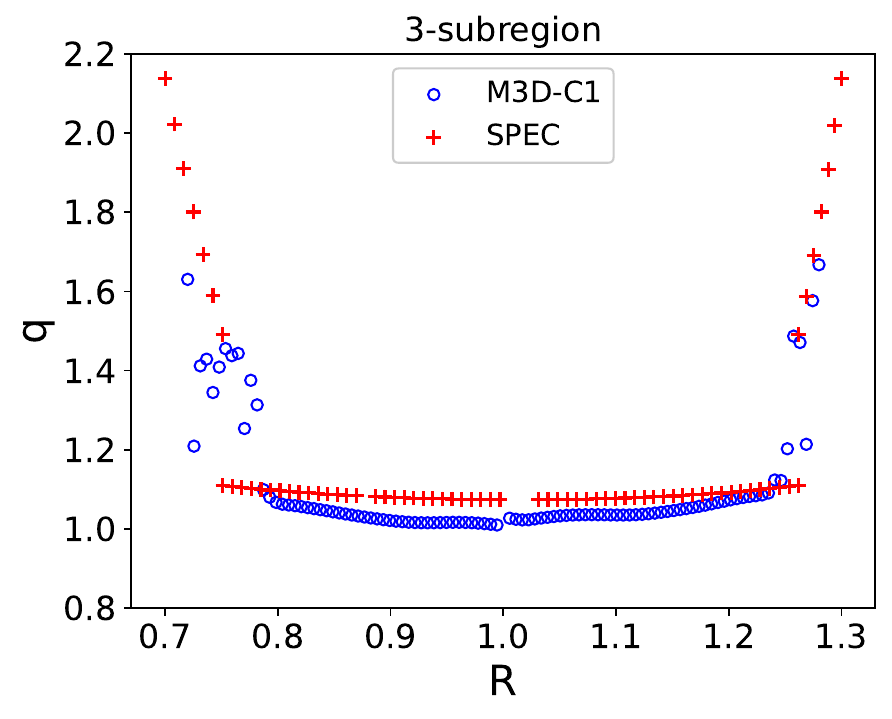}
    \\
    (c) & (d)
    \end{tabular}
    \caption{A comparison between M3D-C1 and SPEC for the safety factor $q$ as a function of field-line tracing starting point $R$, for (a) 15-subregion, (b) 11-subregion, (c) 7-subregion, and (d) 3-subregion (after-crash) equilibrium. The $q$ profile is constructed by field-line tracing starting from $Z=0$ and the given $R$, and computed with respect to the original axis, except for the 3-subregion equilibrium, which is computed with respect to the new axis.
    Some irregularities are found in (c) and (d) due to chaotic region.}
    \label{fig:q_compare}
\end{figure*}

We compare the $q$ profile from both codes, shown in \figref{fig:q_compare}.
The $q$ profile is computed with respect to the original axis by field-line tracing for 100 toroidal turns starting from $Z=0$ and a given $R$, except that in the after-crash case, it is computed with respect to the new magnetic axis.
Overall, the agreement between M3D-C1 and SPEC is remarkable.
For \figref{fig:q_compare} (a) and (b), the $q$ profile outside the reconnection region generally follows the initial configuration.
A flattened $q=1$ region appears around $R=0.85$, corresponding to the O-point of the $m=n=1$ island.
On both side of the $q=1$ region there are jumps and sharp changes in $q$, corresponding to the separatrix of the island.
Around $R=1.23$, there is a jump in $q$ due to the ribbon of the island, indicating the existence of a current-sheet.
We note that SPEC agrees very well with M3D-C1 for both the sharp changes around the island and the jump of $q$ across the ribbon, indicating that the magnitude of the current sheet is correctly captured.
This current sheet will present as an interface current leading to the shearing of $\bm{B}$ across the interface.
Outside the island, the $q$ profile of M3D-C1 is slightly lower than SPEC.
Since M3D-C1 is a global initial value code calculating the total field instead of the perturbation, even though the initial equilibrium is given by the VMEC equilibrium, it evolves according to the set source and resistivity.
For \figref{fig:q_compare} (c), the island becomes much wider and chaotic regions start to develop outside the unreconnected flux surfaces and the new island.
The computed M3D-C1 $q$ value in this region fluctuates due to the chaotic and ergodic nature of the field lines, though the match with SPEC is still relatively good.
Finally, both SPEC and M3D-C1 predict a pretty flat $q$ profile above unity in the core of the after-crash scenario with a similar value.
We note that the sawtooth we are studying is a complete and global crash, which allows us to model the reconnected region as a single Taylor-relaxed region and explains the good match.

\begin{figure*}[!htbp]
    \centering
    \begin{tabular}{c c}
    \includegraphics[width=0.48\linewidth]{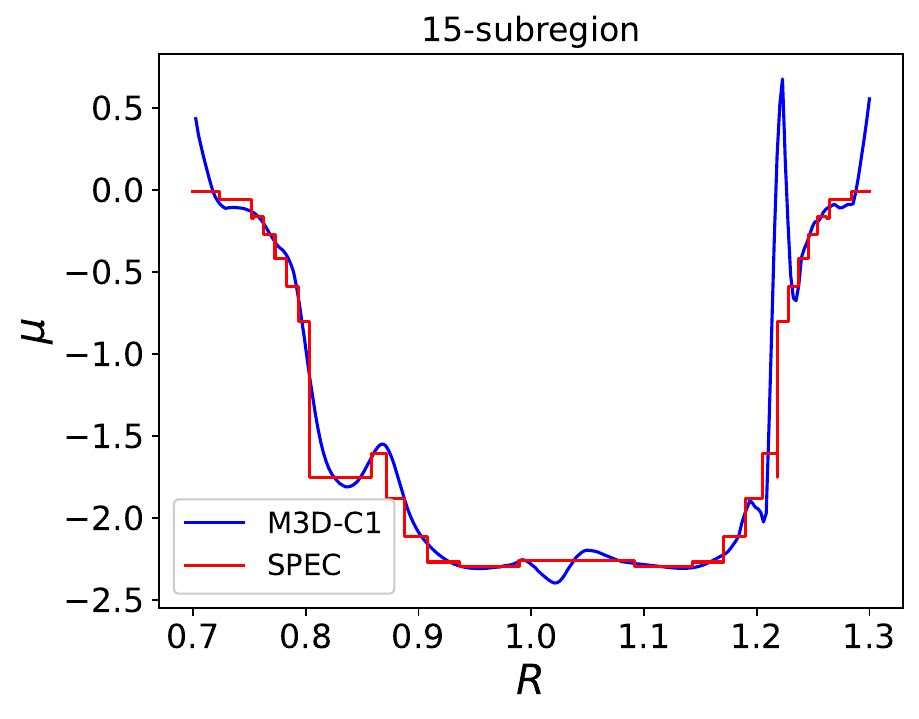}
    &
    \includegraphics[width=0.48\linewidth]{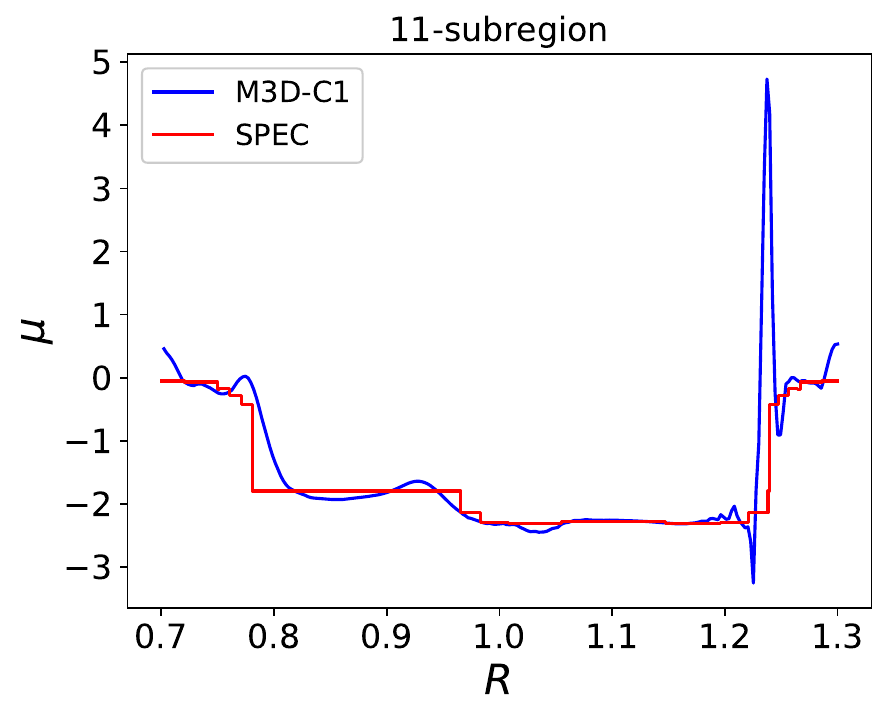}
    \\
    (a) & (b) \\
    \includegraphics[width=0.48\linewidth]{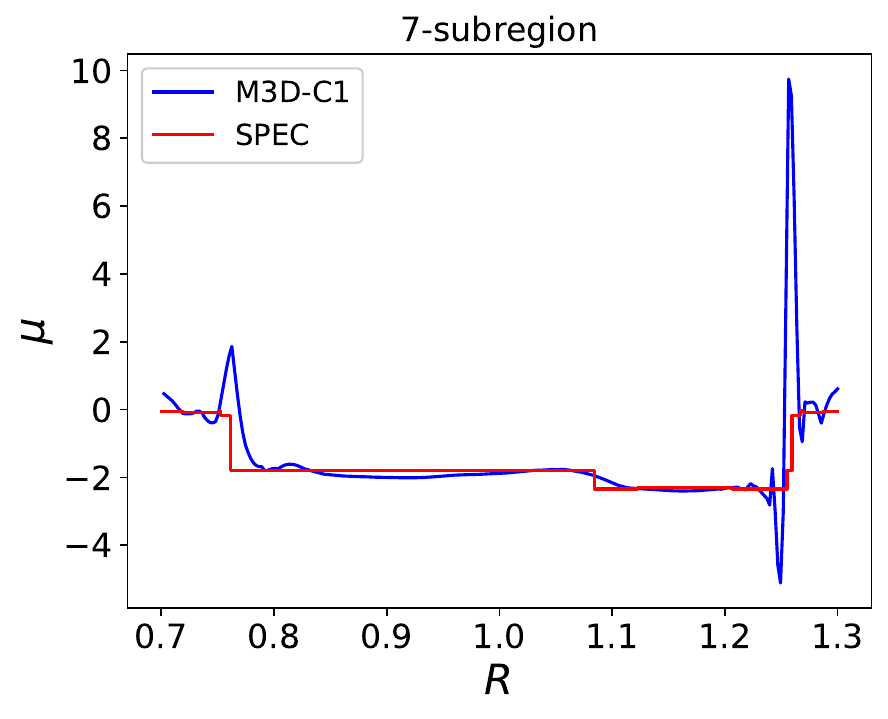}
    &
    \includegraphics[width=0.48\linewidth]{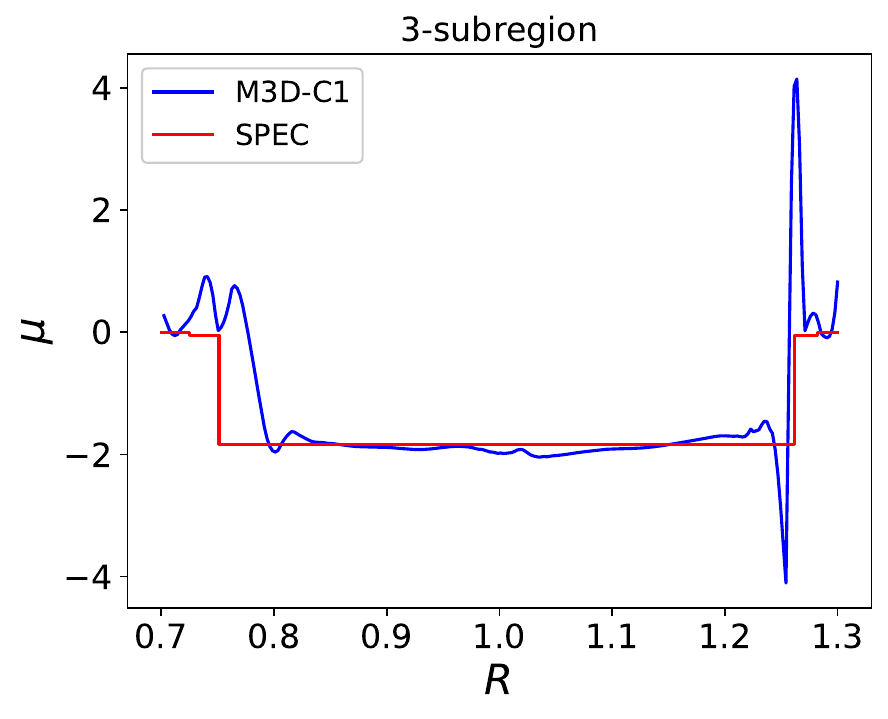}
    \\
    (c) & (d)
    \end{tabular}
    \caption{A comparison between M3D-C1 and SPEC for the profile $\mu=\mu_0 J_\parallel /B$ on the mid-plane, for (a) 15-subregion, (b) 11-subregion, (c) 7-subregion, and (d) 3-subregion (after-crash) equilibrium. For SPEC, the currents flowing on the interfaces are ignored: only currents inside the subregions are considered.}
    \label{fig:mu_compare}
\end{figure*}

In the Kadomtsev model, each pair of flux surfaces with the same helical flux label reconnects into a new, crescent flux surface surrounding the axis of the island.
The quantity $\mu=J_\parallel/B$ is not a constant across these new flux surfaces inside the island, while in SPEC the entire reconnected island region is modelled as a single relaxed volume with a constant $\mu$.
The value of $\mu$ is only allowed to jump across interfaces, leading to a stepped $\mu$ profile.
We examine the appropriateness of this assumption by comparing the value $\mu$ on the mid-plane of M3D-C1 and SPEC, shown in \figref{fig:mu_compare}.
The general agreement between M3D-C1 is again remarkable across all cases despite the discrete nature of $\mu$ in SPEC.
More specifically, the value of $\mu$ appears flat within the reconnected region (around $R=0.82, 0.87, 0.9, 1.0$, for (a-d), respectively) for M3D-C1, matching the value predicted by SPEC.
This justifies our single-subregion assumption for the reconnected region.

\begin{figure}[!htbp]
    \centering
    \includegraphics[width=0.48\linewidth]{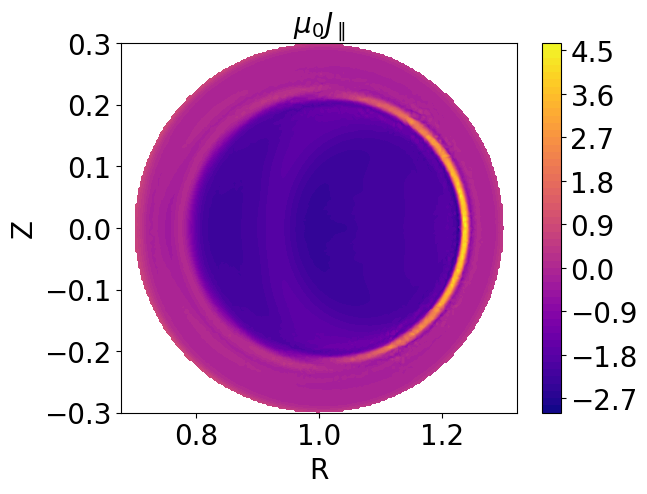}
    \caption{The contour of the parallel current for M3D-C1 time slice (b), corresponding to the 11-subregion SPEC equilibrium. A current sheet is visible in bright colour.}
    \label{fig:Jpar}
\end{figure}

\begin{figure}[!htbp]
    \centering
    \includegraphics[width=0.48\linewidth]{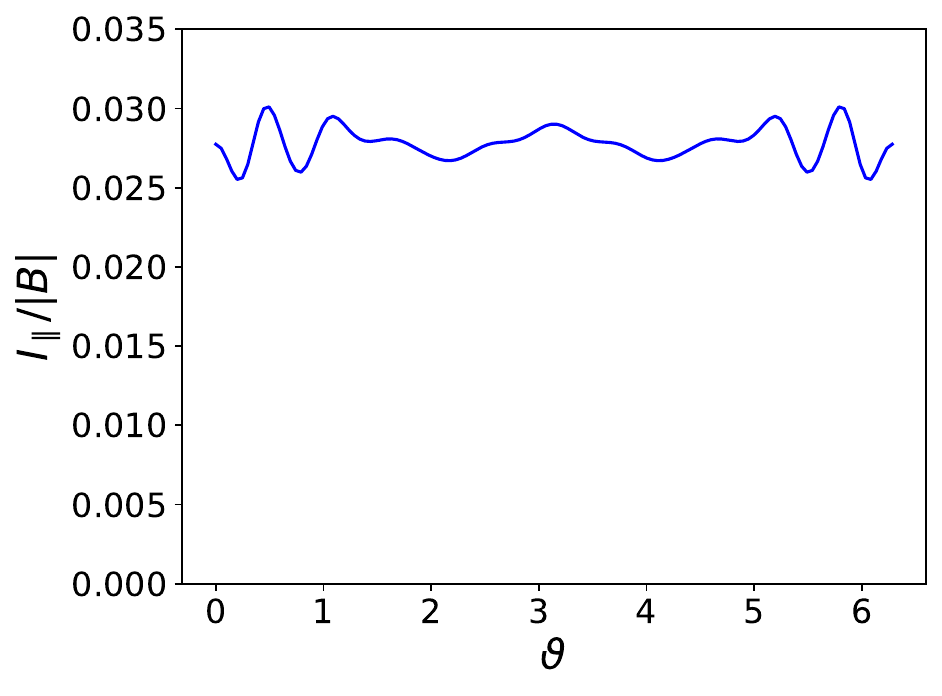}
    \caption{The ratio between the toroidal surface current $I_\parallel$ and field strength $|B|$ as a function of SPEC generalised poloidal angle $\vartheta$ on the interface just outside of the reconnection site for the 11-subregion SPEC equilibrium.}
    \label{fig:surface_I}
\end{figure}

A notable feature of M3D-C1 is the strong peak of current between $R=1.2$ and $1.3$ as a result of the current sheet at the reconnection site.
This current is missing for the SPEC curve since we are only plotting parallel currents flowing within the SPEC subregions and leaving out the current sheet on interfaces.
To compare the structure of the current sheet, we plot the parallel current on a two-dimensional R-Z plane for the M3D-C1 case (b) in \figref{fig:Jpar}.
\figref{fig:Jpar} shows that the current sheet is concentrated at the reconnection site, and is poloidally asymmetric:
its strength decreases away from the middle of the ribbon and weakens significantly outside the ribbon.
We have also computed the SPEC surface current on the interface immediately outside the reconnected volume, defined by~\cite{baillod_computation_2021}
\begin{equation}
    I^\zeta = \frac{1}{J} \left[\left[ B_\vartheta \right]\right], \quad
    I^\vartheta = -\frac{1}{J} \left[\left[ B_\zeta \right]\right]
    \label{eq:surface_current}
\end{equation}
where the superscripts and subscripts indicate the contra-variant and co-variant components, respectively, and $J$ the jacobian.
The parallel surface current $I_\parallel$ is then computed by taking the dot product of the surface currents in \eqref{eq:surface_current} and $\bm{B}$, making use of the metric.
Note that to resolve the ambiguity of $\bm{B}$ and metric on different sides of the interface, we use their value on the inner side of the interface.
In \figref{fig:surface_I}, we plot the ratio $I_\parallel/|B|$ as a function of SPEC generalised poloidal angle $\vartheta$, which appears to be almost a constant.
We therefore conclude that a SPEC equilibrium cannot resolve the poloidal structure of the current sheet.
Ideally, one can integrate within a neighbourhood of the M3D-C1 current sheet to get the poloidally averaged surface current amplitude and compare it to SPEC.
However, its value is very sensitive to the range of integration and a direct comparison with SPEC is difficult.
Nevertheless, the average amplitude of the current sheet is correct, which can be reflected by the same jump of the safety factor across the current sheet as shown in \figref{fig:q_compare}.

Finally, we calculate the drop in magnetic energy in M3D-C1 and have overplotted it in \figref{fig:energy}, taking the time slice $t=1350$, just before the crash as a reference.
Compared to SPEC, the magnetic energy in M3D-C1 was reduced more in the early stage.
This is because, in M3D-C1, the entire plasma is allowed to respond resistively to the crash, while in SPEC, only the inner region containing the reconnection site is relaxed and the rest of the plasma volumes are constrained by ideal interfaces and fixed helicity/fluxes.
At the later stage, as the relaxation volume increases and the number of subregions reduces in SPEC, the match becomes better.

\section{Conclusion}
\label{sec:conclusion}
In this study, we have constructed a sequence of Multi-region Relaxed MHD (MRxMHD) equilibria to represent the sawtooth crash process in tokamak plasmas with zero plasma $\beta$. 
The plasma is partitioned into subregions separated by ideal interfaces, allowing the relaxation locally within reconnected regions while preserving the magnetic topology outside. 
A helical equilibrium with a magnetic island was identified as having lower energy than the corresponding axisymmetric state. 
The reconnection region was then progressively enlarged by removing interfaces in pairs, representing the intermediate stages of the sawtooth crash, with the island growing larger and the magnetic energy further reducing.
These helical equilibria are all minimum energy states given the constrained topology of the unreconnected region, with the two interfaces bounding the island barely touch each other:
the only way to reduce the energy further is via reconnecting these two interfaces.
The results demonstrate the capability of MRxMHD in capturing the transition from the initial axisymmetric equilibrium to states with magnetic islands, chaotic regions, and, eventually, the post-crash equilibrium.
\modi{The MRxMHD approach does not of course capture the time dependence of the transition.}

Comparison with nonlinear MHD simulations using M3D-C1 reveals qualitative and quantitative agreement: this confirms the sawtooth with low $\beta$ considered in our current paper is indeed a reconnection process.
\modi{
We demonstrated that each state along the path of reconnection is a minimum energy state with the reconnected region being (nearly) Taylor-relaxed and the crescent flux surfaces self-emerging.
}
The jump of the $q$ profile matches across the separatrix, 
showing that MRxMHD is correctly capturing the size of the current sheet.
The MRxMHD approach fails to spatially-resolve the poloidal structure of the current sheet due to its coarse assumption of the reconnected region being a single Taylor-relaxed subregion,
despite the flat current profile being a good assumption for the majority of the island.
\modi{Combined, these results demonstrate that MRxMHD approach is able to model the path of reconnection, not just the post-crash state.}
\modii{The construction of SPEC equilibria from a given VMEC equilibrium is fully automated once the number of subregions is specified. Each SPEC equilibrium typically requires approximately 5 hours of wall-clock time, using a number of CPUs equal to the number of subregions. For comparison, the M3D-C1 simulation presented in this work required a total of 5,000 CPU hours. This highlights the computational efficiency of the MRxMHD approach, which enables the direct construction of intermediate states during the crash.}  

Future work may focus on extending this approach to finite-$\beta$ plasmas.
At a higher $\beta$, the property of the instability might become qualitatively different (e.g. the potential transition to the Wesson model).
It is worthwhile to explore such a transition and see what MRxMHD predicts as the minimum energy states.
We also plan to explore its application in more realistic tokamak scenarios and in stellarators such as the ECCD-induced crash in W7-X.
Finally, the new relaxed MHD framework developed by Dewar~\cite{dewar_relaxed_2022} recently might give more insights since it contains plasma flow and allows time evolution.

\ 

\begin{acknowledgments}
We dedicate this article to the late great Em. Prof. Robert (Bob) Dewar, who founded the MRxMHD theory, and put forward the initial idea of this work.
We acknowledge Dr. Graham Dennis for his preliminary work during his time at the ANU.
We also thank the colleagues from the Simons Collaboration on Hidden Symmetries and Fusion Energy for fruitful discussions. 

Z.S.Q. is supported by National Research Foundation Singapore (NRF) project ``fusion science for clean energy'', and Ministry of Education (MOE) AcRF Tier 1 grants RS02/23 and RG156/23.
Y.Z. was supported by the National Natural Science Foundation of China under Grant Number 12305246 and the Fundamental Research Funds for the Central Universities.
This work was supported by a grant from the Simons Foundation/SFARI (560651, AB; and 1013657, JL). This work has been carried out within
the framework of the EUROfusion Consortium, via the Euratom Research and Training Programme (Grant Agreement No. 101052200—EUROfusion) and funded by the Swiss State
Secretariat for Education, Research and Innovation (SERI). Views and opinions expressed are however those of the author(s) only and do not necessarily reflect those of the European Union, the European Commission, or SERI. Neither the European Union nor the European Commission nor SERI can be held responsible for them.

The computational work for this article was partially performed on resources of the National Supercomputing Centre (NSCC), Singapore.
This research was undertaken partially with the use of the National Computational Infrastructure (NCI Australia). NCI Australia is enabled by the National Collaborative Research Infrastructure Strategy (NCRIS).
\end{acknowledgments}

\section*{Data Availability Statement}

The data that support the findings of this study are available from the corresponding author upon reasonable request.

\appendix


\bibliography{ref}

\end{document}